\documentclass[10pt,aps,prd,superscriptaddress,nofootinbib,nobibnotes,longbibliography,floatfix,twocolumn]{revtex4-2}
\usepackage{float}
\usepackage{bm}
\usepackage{mathtools,
amsmath,
amssymb,
amsfonts,
mathrsfs,
chngcntr,
multirow}

\let\cc\corresponds
\let\corresponds\relax
\usepackage{mathabx}
\let\corresponds\cc

\usepackage[utf8]{inputenc}
\usepackage[T1]{fontenc}

\usepackage[dvipsnames]{xcolor}
\usepackage[unicode]{hyperref}
\hypersetup{colorlinks=true, citecolor=MidnightBlue,
            linkcolor=MidnightBlue, urlcolor=MidnightBlue, linktocpage=true}
\usepackage[normalem]{ulem}

\DeclareMathAlphabet{\mathpzc}{OT1}{pzc}{m}{it}
\definecolor{darkgreen}{rgb}{0.0, 0.6, 0.0}

\newcommand{\ie}{i.e.}

\newcommand{\note}[1]{\text{\scshape\tiny{#1}}}

\newcommand{\ii}{\mathrm{i}}
\newcommand{\dd}{\mathrm{d}}

\newcommand{\Ord}{\mathcal{O}}

\newcommand{\al}{\alpha}
\newcommand{\be}{\beta}

\newcommand{\de}{\delta}
\newcommand{\De}{\Delta}

\newcommand{\la}{\lambda}

\newcommand{\om}{\omega}

\newcommand{\dl}{\partial}

\usepackage{orcidlink}

\newcommand{\orcid}[1]{\href{https://orcid.org/#1}{\includegraphics[width=10pt]{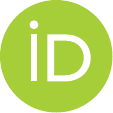}}}

\begin{document}

\title{Quasinormal modes of rotating black holes beyond general relativity \\in the WKB approximation}

\author{Ruijing Tang \orcid{0009-0000-2737-4078}}
\email{Ruijing.TANG@obspm.fr}
\affiliation{Observatoire de Paris, Universite PSL, 61 avenue de l'Observatoire, 75014 Paris, France}

\author{Nicola Franchini \orcid{0000-0002-9939-733X}}
\email{nicola.franchini@tecnico.ulisboa.pt}
\affiliation{CENTRA, Departamento de F\'{\i}sica, Instituto Superior T\'ecnico -- IST, Universidade de Lisboa -- UL, Avenida Rovisco Pais 1, 1049-001 Lisboa, Portugal}

\author{Sebastian H.\,V\"olkel \orcid{0000-0002-9432-7690}}
\email{sebastian.voelkel@aei.mpg.de}
\affiliation{Max Planck Institute for Gravitational Physics (Albert Einstein Institute), \\ Am M\"uhlenberg 1, 
D-14476 Potsdam, Germany}

\author{Emanuele Berti \orcid{0000-0003-0751-5130}}
\email{berti@jhu.edu}
\affiliation{William H. Miller III Department of Physics and Astronomy, Johns Hopkins University, Baltimore, Maryland 21218, USA}

\begin{abstract}
Exploring gravitational theories beyond general relativity (GR) with black hole (BH) spectroscopy requires accurate and flexible methods for computing their quasinormal mode (QNM) spectrum. 
A popular method of choice is the higher-order Wentzel–Kramers–Brillouin (WKB) approximation, mostly applied to nonrotating BHs. 
While previous studies demonstrated that the higher-order WKB method can also be used for Kerr BHs in GR, there has been little work on  rotating BHs in modified theories of gravity. 
In this work, we revive the idea by extending WKB calculations of the Kerr QNM spectrum to higher order and assessing its accuracy against continued-fraction tabulated data. 
We then apply the WKB approximation beyond GR, comparing it against both linearized and continued fraction calculations in the parametrized beyond-Teukolsky formalism and in higher-derivative gravity theories. 
We find that the frequencies computed by the WKB method in theories beyond GR have better accuracy than the measurement errors for GW250114, the event with the highest ringdown signal-to-noise ratio observed to date.
\end{abstract}

\maketitle

\section{Introduction}

The observation of gravitational waves from compact binary mergers by the LIGO-Virgo-KAGRA (LVK) Collaboration in the past decade has revolutionized gravitational physics and astronomy~\cite{LIGOScientific:2016aoc,LIGOScientific:2017vwq,LIGOScientific:2018mvr,KAGRA:2021vkt,LIGOScientific:2020ibl,LIGOScientific:2020ibl}. 
In particular, it allows us to test the strong and dynamical regime of general relativity (GR) and the nature of black holes (BHs)~\cite{LIGOScientific:2016lio,LIGOScientific:2019fpa,LIGOScientific:2020tif,LIGOScientific:2021sio}. 

The standard assumption is that binary BH mergers produce a perturbed Kerr BH that settles into an equilibrium configuration via the emission of gravitational waves. 
Long before numerical relativity simulations were able to simulate the complete binary merger and ringdown process~\cite{Pretorius:2005gq,Buonanno:2006ui,Berti:2007fi}, linear perturbation theory predicted that part of the waveform can be expanded in terms of characteristic quasinormal modes (QNMs). 
In GR, the linear axial and polar metric perturbations of Schwarzschild BHs are governed by the Regge-Wheeler and Zerilli equations, respectively~\cite{Zerilli:1970se,Regge:1957td}. 
The more general case of Kerr BHs is governed by the Teukolsky equation, derived in the Newman-Penrose formalism, where the perturbed quantities refer to the complex Weyl scalars~\cite{Teukolsky:1972my,Teukolsky:1973ha}. 
These so-called master equations reduce the full problem to the solution of radial wave equations with certain effective potentials. In the case of the Teukolsky equation, we have to simultaneously solve a nontrivial angular equation. 
The QNMs form a discrete set of complex eigenvalues for each angular multipole $(\ell,\,m)$, and they are normally sorted through an ``overtone number'' $n=0,\,1,\,2\dots$ in terms of the magnitude of their imaginary part. The QNMs are solutions of the master equations corresponding to physically motivated boundary conditions: purely outgoing waves at spatial infinity and purely ingoing waves at the horizon. 

In GR and within the assumptions of the Kerr hypothesis~\cite{Kerr:1963ud,Carter:1971zc,Robinson:1975bv}, QNMs depend only on the Kerr BH mass and spin. 
By measuring more than one QNM one can test GR and the Kerr hypothesis by performing ``black hole spectroscopy''~\cite{Detweiler:1980gk,Echeverria:1989hg,Finn:1992wt,Dreyer:2003bv,Berti:2005ys}. 
We refer the reader to Refs.~\cite{Kokkotas:1999bd,Nollert:1999ji,Berti:2009kk,Konoplya:2011qq} for early reviews on QNMs, Ref.~\cite{Berti:2018vdi,Franchini:2023eda} for more specialized reviews on ringdown tests of GR, Ref.~\cite{Carullo:2025oms} for an observational status update, and Ref.~\cite{Berti:2025hly} for the most recent and detailed review article. 

The computation of QNMs for rotating BHs beyond GR is one of the grand challenges in gravitational wave research. 
Even if one succeeds in the challenging task of using perturbation theory to derive a set of decoupled master equations, it is still far from trivial to obtain the corresponding QNM spectrum.  
Complicated master equations, like those arising for rotating BHs beyond GR, limit the list of tools that are commonly available in the much simpler Schwarzschild case. 
A first complication is that the effective potential and separation constant depend on the complex valued QNMs, and therefore the angular and radial equations must be solved simultaneously. 
Methods that work in GR, e.g., based on the continued fraction relations first employed by Leaver~\cite{Leaver:1985ax}, may not always be useful. 
Recent ``brute force'' approaches based on spectral methods can be used to compute BH QNMs even if the perturbations are not separable~\cite{Chung:2023wkd,Chung:2024vaf,Blazquez-Salcedo:2023hwg,Blazquez-Salcedo:2024oek,Khoo:2024agm,Blazquez-Salcedo:2024dur}. 
Despite this significant advantage, these methods are technically challenging and computationally  expensive. 

A popular and versatile method to compute BH QNMs is based on Wentzel–Kramers–Brillouin (WKB) theory~\cite{Bender:1999box}. 
The WKB approximation can be applied to typical BH potentials, that display a single barrier close to the light ring with two classical turning points.  
The theory has been systematically extended to higher orders~\cite{Schutz:1985km,Iyer:1986np,Iyer:1986nq,Kokkotas:1988fm,Konoplya:2003ii,Matyjasek:2017psv}. 
Within the range of its applicability (that we discuss in more detail below), it provides the QNM spectrum for low-lying QNMs ($n \lesssim \ell$) directly in terms of the coefficients of a Taylor expansion of the effective potential around its maximum. 
This simple recipe works even when the effective potential is only known numerically~\cite{Suvorov:2021amy}. 
This advantage is crucial if the modified perturbation equations are not amenable to any other standard methods (for example because, as in the case of Leaver's method~\cite{Leaver:1985ax}, one requires analytic expansions). 

Despite its widespread use for nonrotating BHs, applications of the WKB approximation to rotating BHs in GR and beyond are relatively rare. 
It was first applied to gravitational perturbations of Kerr BHs in Refs.~\cite{Seidel:1989bp,Kokkotas:1991vz}, and later applied to scalar field perturbations around rotating regular BHs and wormholes in Ref.~\cite{Franzin:2022iai}. 
Note that a closely related application of WKB theory is the so-called eikonal (large $\ell$) approximation. 
In this context, WKB theory has been applied to Kerr BHs in Ref.~\cite{Yang:2012he}, and used to study QNMs beyond GR in Refs.~\cite{Glampedakis:2017dvb,Glampedakis:2019dqh,Silva:2019scu,Bryant:2021xdh}. 
Note however that the higher-order WKB method is, in principle, an accurate tool to obtain the QNMs for a potential barrier, while the eikonal approximation used to simplify the structure of the perturbation equations is typically of limited accuracy for small $\ell$. 
Moreover, for some beyond-GR theories, one cannot always link eikonal QNMs to unstable circular photon orbits~\cite{Konoplya:2017wot}. 
A WKB analysis that avoids the expansion used in the higher-order WKB method has been presented in Ref.~\cite{Miyachi:2025ptm}. 
The eikonal approximation was also used to study the QNMs for rapidly rotating black holes in higher-curvature gravity in Refs.~\cite{Cano:2025mht,Cano:2025ejw}, which provide promising results for testing even small deviations from GR.
 
In this work, we first carry out a comprehensive application of the higher-order WKB method to Kerr BHs in order to quantify the method's range of applicability.  
Then we report the first application of the higher-order WKB method to rotating BHs beyond GR, carrying out both theory-agnostic and theory-specific calculations. 
To represent the large class of possible deviations, we work within the recently developed parametrized beyond-Teukolsky QNM framework~\cite{Cano:2024jkd}, which can be seen as an extension of the parametrized QNM framework beyond Schwarzschild~\cite{Cardoso:2019mqo,McManus:2019ulj,Volkel:2022aca,Volkel:2022khh,Franchini:2022axs,Kimura:2020mrh,Hirano:2024fgp}. 
The key assumptions underlying the beyond-Teukolsky formalism are that modifications to GR can be captured by linear corrections, and that the master equations are separable. 
We use the formalism as a test bed to study the reliability of the higher-order WKB method. 
Note that the WKB method itself is not limited to small deviations from GR, but instead restricted to the usual assumptions underlying WKB theory, as well as the assumption that the effective potential has a potential barrier (see e.g.~\cite{Konoplya:2019hlu} for a discussion).

The paper is organized as follows. 
In Sec.~\ref{sec:GR} we review the basics of the higher-order WKB method and of its application to the Teukolsky equation in GR. 
In Sec.~\ref{sec:bGR} we discuss the modified Teukolsky equation and extend the WKB method to this case. 
In Secs.~\ref{sec:res_GR} and~\ref{sec:res_bGR}
we report our main results for the GR case and for the modified Teukolsky equation, respectively. 
Our conclusions can be found in Sec.~\ref{sec:conc}. 
As is common in BH perturbation theory, we use units such that $G=c=1$ and we set the BH mass $M=1$.

\section{WKB method for rotating black holes}\label{sec:GR}

In this section, we briefly review the higher-order WKB method and its application to the Teukolsky equation. 

\subsection{Theoretical minimum of the WKB method}\label{sec:WKB_theory}

Let us start from a perturbation equation of the form
\begin{equation}\label{eq:canonical_equation}
    \frac{\dd^2 R}{\dd r_*^2} + Q(\om,B;r_*) R = 0 \,,
\end{equation}
where $\om$ are the eigenfrequencies, $B$ is the separation constant of the equation, and $r_*$ is the tortoise coordinate. 
To define the contributions to the higher-order WKB formula, we introduce the $N$th derivative of the potential evaluated at the peak
\begin{equation}
    Q_N \equiv \left.\frac{\dd^N Q}{\dd r_*^N}\right|_{\overline{r}_*} \,,
\end{equation}
where the peak radius $\overline{r}_*$ is found by solving the equation
\begin{equation}\label{eq:peak}
    Q_1(\om,\overline{r}_*) = 0 \,.
\end{equation}
In the case of nonrotating Schwarzschild BHs, or similar cases to which the method is usually applied, this simply corresponds to finding the maximum of the effective potential. 
This maximum is real valued and close to the location of the light ring. 
In the more general case studied here, the ``maximum'' is complex valued and requires a more careful treatment~\cite{Seidel:1989bp,Kokkotas:1991vz,Franzin:2022iai}. 

In general, one has that the $N$th order WKB formula contains up to the 2$N$th derivative of $Q$ at the peak
\begin{equation}\label{eq:WKB_Nth}
    F_N \equiv F_N[Q_0,Q_2,Q_3,...,Q_{2N}] = 0 \,.
\end{equation}
The explicit expression of $F_N$ can be found in the literature~\cite{Iyer:1986nq,Konoplya:2019hlu,Matyjasek:2017psv}. 
Here, for illustration, we report the first-order WKB formula
\begin{equation}\label{eq:WKB_first}
   F_1[\om,\overline{r}_*] \equiv Q_0 + \ii \be \sqrt{2 Q_2} = 0 \,,
\end{equation}
where $\be = n + 1/2$, and $n$ is the overtone number. 
This agrees with the Schutz-Will formula, which approximates the effective potential with a parabola and determines the QNMs from asymptotic matching of the parabolic cylinder functions~\cite{Schutz:1985km}. 
Note that the underlying WKB series is an asymptotic series, and it is not guaranteed that higher orders will give more accurate results. 
In the context of BH QNMs, one usually finds that higher-order WKB corrections are particularly important for improving the accuracy of overtones~\cite{Iyer:1986np}.

\subsection{The WKB method and the Teukolsky equation}\label{WKB rotating BH}

Linear perturbations of the Kerr BH are described by the Teukolsky equation. 
The radial Teukolsky equation~\cite{Teukolsky:1972my,Teukolsky:1973ha} for a spin-$s$ perturbation reads
\begin{equation}
    \frac{1}{\De^s} \frac{\dd}{\dd r}\left[ \De^{s+1} \frac{\dd R(r)}{\dd r}\right] + V(r)R(r) = 0\,,
\end{equation}
where the effective potential is
\begin{equation}
    V(r) = 2\ii s \frac{\dd K}{\dd r} - \la_{\ell m} + \frac{1}{\De}\left( K^2 -\ii s K \frac{\dd \De}{\dd r} \right) \,,
\end{equation}
and we defined
\begin{align}
    \De & \, = r^2 - 2r +a^2\,, \qquad K = (r^2+a^2)\om - a m \,, \\
    \la_{\ell m} & \, = B_{\ell m} + a^2\om^2 -2 a m \om \,.
\end{align}
Here, $\ell$ and $m$ are the angular momentum and azimuthal number of the perturbation, and $B_{\ell m}$ is the separation constant. 
We want to cast the Teukolsky equation in the form of Eq.~\eqref{eq:canonical_equation}. 
To do so, we introduce the new field
\begin{equation} \label{eq:field_redef}
 Y = \Delta^{s/2} \left( r^{2} + a^{2} \right)^{1/2} R \,,
\end{equation}
which satisfies the equation
\begin{equation} \label{eq:teuk_canonical}
    \frac{\dd^2 Y}{\dd r_*^2}+ \left[ V\frac{\Delta}{(r^{2} + a^{2})^{2}} - G^{2} - \frac{\dd G}{\dd r_*} \right] Y = 0 \,,
\end{equation}
where we have introduced the tortoise coordinate
\begin{equation}
\quad \frac{dr_*}{dr} = \frac{r^{2} + a^{2}}{\Delta} \,,
\end{equation}
and
\begin{equation}
G = \frac{s(r-1)}{r^2+a^2} + \frac{r \De}{\left( r^2 + a^2 \right)^2}\,.    
\end{equation}
From Eq.~\eqref{eq:canonical_equation}, we can define
\begin{equation}
    Q^\note{GR} = V\frac{\Delta}{(r^{2} + a^{2})^{2}} - G^{2} - \frac{\dd G}{\dd r_*} \,.
\end{equation}
The computation of the QNMs of the Kerr metric with the WKB method proceeds as follows. First of all, we specialize to gravitational perturbations ($s = -2$).
Then, we start applying the WKB method from $a=0$. In this case, the separation constant reduces to $B = \ell(\ell+1) -2$, and Eq.~\eqref{eq:peak} for the location of the peak reduces to
\begin{equation}
    \left(1 - \frac{2}{r}\right) \left[ \frac{2\ii\om}{r^2} + \frac{B+2-12\ii\om}{r^3} - 3\frac{B+1}{r^4} \right] = 0 \,,
\end{equation}
whose nontrivial solutions are
\begin{equation}\label{eq:max_r}
    \overline{r}_\pm = 3 - \la \pm \sqrt{\la^2 - \frac{3}{2\ii\om} + 9} \,,
\end{equation}
with $\la = \ell(\ell+1)/4\ii\om$. 
We have checked numerically that picking $\overline{r}_+$ provides the correct values for Schwarzschild QNMs.
Now, we can solve the equation 
\begin{equation}
    F_N[\om,\overline{r}_+;Q^\note{GR}(a=0)] = 0
\end{equation}
using a numerical root-finding method to obtain the estimate of the eigenfrequency $\om_N$ at the $N$th WKB order.
Once we have $\overline{r}_+$ and $\om_N$ at $a=0$, we can increase the spin to solve Eqs.~\eqref{eq:peak} and \eqref{eq:WKB_Nth} simultaneously for $\overline{r}$ and
$\om$ with a numerical root-finding scheme, using the values at the previous step in the spin as initial guesses.

\section{WKB and modified Teukolsky equation}\label{sec:bGR}

\subsection{Modified Teukolsky equation}

We assume that the Teukolsky equation is modified as
\begin{equation}\label{eq:mod_teukolsky}
    \frac{1}{\De^s} \frac{\dd}{\dd r}\left[ \De^{s+1} \frac{\dd R(r)}{\dd r}\right] + V(r)R(r) + \de V(r) R(r) = 0\,.
\end{equation}
In general, the modification $\de V(r)$ can take any form. 
Modified Teukolsky-type equations in beyond-GR settings were derived in~\cite{Li:2022pcy,Hussain:2022ins,Cano:2023tmv}. 
The parametrized QNM framework for the modified Teukolsky equation~\cite{Cano:2024jkd} has recently been introduced as a conceptual extension of the parametrized QNM framework for Schwarzschild BHs~\cite{Cardoso:2019mqo,McManus:2019ulj,Volkel:2022aca,Volkel:2022khh,Hirano:2024fgp,Franchini:2022axs,Kimura:2020mrh,Hirano:2024fgp}. 
The explicit form of the modifications in powers of $r$ is given by
\begin{equation}
\label{eq:deltav-fuc1}
    \de V(r) = \sum_{k=-K}^{4} \al^{(k)} \de V_k(r) \,,
\end{equation}
where
\begin{equation}
\label{eq:deltav-fuc2}
    \de V_k(r) = \frac{1}{\De} \left( \frac{r}{r_+} \right)^k \,.
\end{equation}
To apply the WKB method, we need to recast the equation into the form of Eq.~\eqref{eq:canonical_equation}. By performing the same transformation as in Eq.~\eqref{eq:field_redef}, we obtain
\begin{equation} \label{eq:bteuk_canonical}
    \frac{\dd^2 Y}{\dd r_*^2}+ \left(Q^\note{GR} + \sum_k \al^{(k)} \de Q_k\right) Y = 0 \,,
\end{equation}
where we identified
\begin{equation}
    \de Q_k \equiv \de V_k \frac{\Delta}{(r^2 + a^2)^2} \,.
\end{equation}
In principle, Eq.~\eqref{eq:bteuk_canonical} is ready for the application of the WKB method following the steps outlined in Sec.~\ref{sec:GR}. Before analyzing applications of the WKB method to theories beyond GR, we outline a significant simplification in the limit of small perturbations.

\subsection{Linearized WKB beyond Teukolsky \label{lizwkb}}

In general, the higher-order WKB method can be applied to any beyond-GR effective potential, as long as the underlying WKB assumptions are satisfied. 
Perturbations of nonrotating BHs usually yield a frequency-independent, real-valued potential with two turning points that vanishes asymptotically, so these conditions are satisfied, but counterexamples are known~\cite{Konoplya:2019hlu}. 
Here we consider a potential that has a ``small'' shift with respect to GR, \ie, in the notation of the previous section, the quantities $\al^{(k)}$ are small enough that the linear approximation is sufficient to compute QNMs  (see~\cite{Cano:2024jkd}). 
A similar approach has been applied in the nonrotating case in Ref.~\cite{Volkel:2022khh}. 
In order to make the linear approximation explicit in the definition of the potential, we define
\begin{equation}
    Q = Q^\note{GR} + \al \, \de Q \,,
\end{equation}
where we have dropped the $k$ label for simplicity and the coupling constant $\al \ll 1$.
Such a modification would shift the frequencies and the separation constants by correction terms that are (to leading order) linear in $\al$~\textcolor{red}{\cite{Cano:2024jkd}}:
\begin{align}
    \om = & \om^\note{GR} + \al \, d_\om \,, \label{eq:domega} \\
    B = & B^\note{GR} + \al \, d_B \,. \label{eq:dB}
\end{align}
Similarly, the location of the peak is shifted to $\overline{r}_* = \overline{r}_*^\note{GR} + \al\,\de r_*$. To obtain $\de r_*$, we keep terms linear in $\al$ in Eq.~\eqref{eq:peak}:
\begin{equation}
    Q_1 \simeq Q_1^\note{GR} + \al Q_2^\note{GR} \de r_* + \al \de Q_1 \,.
\end{equation}
Since each term is evaluated on a GR background, the first term on the right-hand side must vanish. 
By requiring the terms linear in $\al$ to vanish as well, we get
\begin{equation}
    \de r_* = - \left. \frac{\de Q_1}{Q_2^\note{GR}} \right|_{\overline{r}_*^\note{GR}} \,.
\end{equation}
Once we determine the new location of the peak, we can parametrize the derivatives of the potential at the peak in terms of $\al$ as follows:
\begin{equation}
\begin{split}
    Q_n &= Q_n^\note{GR} + \al \, \de\overline{Q}_n \,, \\
    \de\overline{Q}_n &= \de {Q}_n +\de r_* Q_{n+1}^\note{GR} \,.
\end{split}
\end{equation}
By plugging everything in Eq.~\eqref{eq:WKB_Nth}, evaluated at the global peak, we have
\begin{equation}
    F_N[Q_i^\note{GR} + \al \de \overline{Q}_i] = 0 \,,
\end{equation}
where $i\in [0,2,\dots,2N]$.
We linearize in $\al$ again to get
\begin{equation}\label{eq:step2}
    \left.F_N\right|_\note{GR} + \left.\al \frac{\dd F_N}{\dd \al}\right|_\note{GR} + \Ord(\al)^2 = 0 \,,
\end{equation}
where each term is evaluated at its GR value $\al =0$, $r_* = \overline{r}_*^\note{GR}$, and $\om = \om^\note{GR}$.
The first term vanishes by definition. By requiring the vanishing of the second term, we get
\begin{equation}
    \left.\frac{\dl F_N}{\dl \al} + \frac{\dl F_N}{\dl \om}\frac{\dl \om}{\dl \al} + \frac{\dl F_N}{\dl B}\frac{\dl B}{\dl \al} \right|_\note{GR} = 0 \,.
\end{equation}
From Eqs.~$\eqref{eq:domega}$ and \eqref{eq:dB} we obtain the $N$th-order WKB estimates for the deviations $d_\om$,
\begin{equation}\label{eq:linear_coeff_WKB}
    d_\om^N = \left. - \frac{1}{\dl F_N /\dl \om} \left( \frac{\dl F_N}{\dl B} d_B + \frac{\dl F_N}{\dl \al} \right) \right|_\note{GR} \,.
\end{equation}
In this equation, we are assuming that the deviations in the angular separation constant are known. As long as the perturbations away from GR are small, we can read off the coefficients $d_B$ obtained in different calculations following the procedure used in Ref.~\cite{Cano:2024jkd}. We employ this assumption for all the beyond-GR deviations studied in this work.

\section{Results for Kerr}\label{sec:res_GR}

\begin{figure*}
    \centering
    \includegraphics[width=\linewidth]{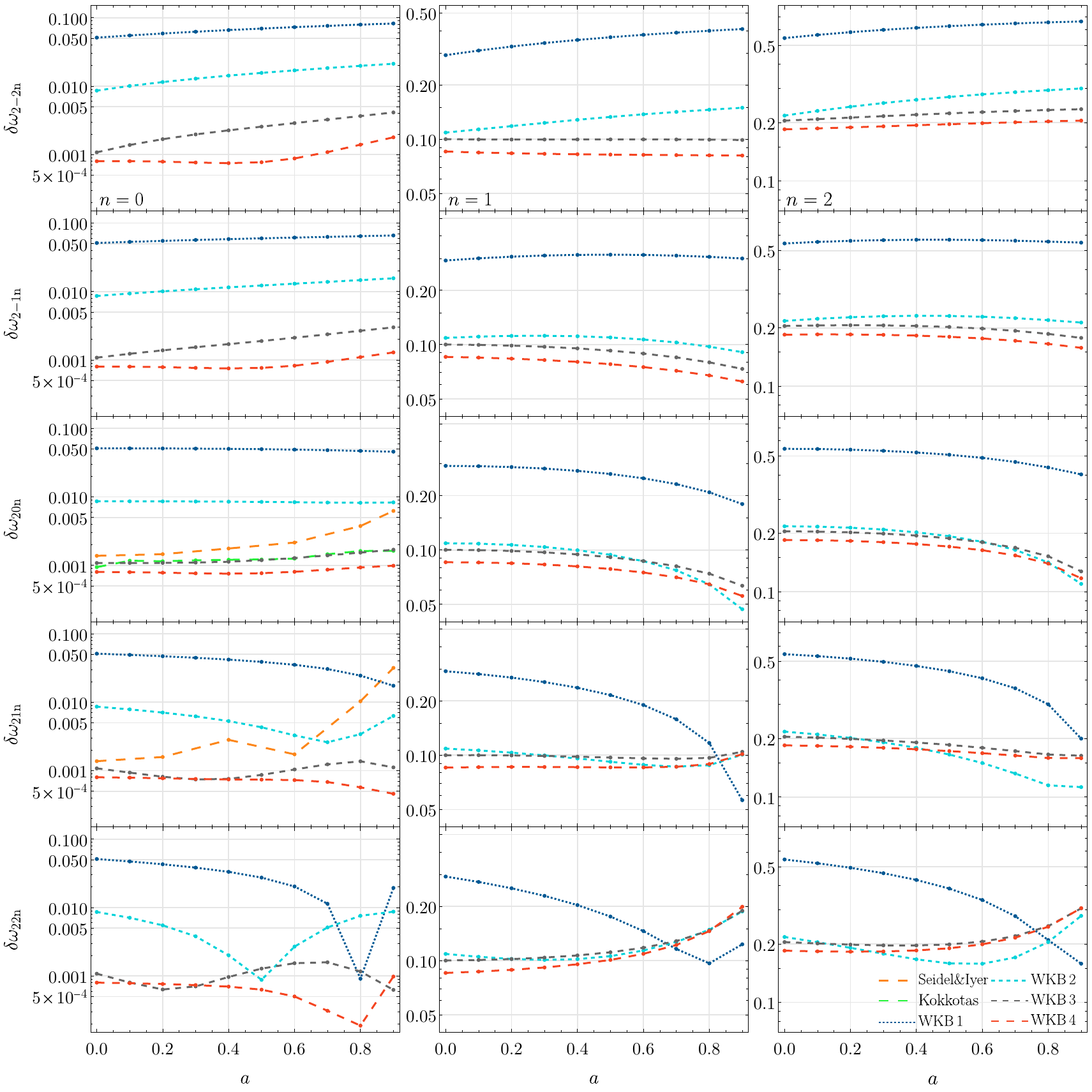}
    \caption{Comparison between WKB approximations (1st to 4th order) and Leaver’s spectral method values for a Kerr BH when changing $(\ell,m,n)$. The vertical axis shows the relative error $\delta \omega _{\ell m n}$ as defined in Eq.~\eqref{rel_err_kerr}, while the horizontal axis spans the spin parameter $a$ from 0 to 0.9. Color indicates WKB orders: first order (dark blue: WKB1), second order (light blue: WKB2), third order (gray: WKB3),  fourth order (light red: WKB4), and Kokkotas' WKB at third order~\cite{Kokkotas:1991vz} (green).
\label{fig:kerrgrwkb}}
\end{figure*}

\begin{figure*}[ht!]
    \centering
    \includegraphics[width=\linewidth]{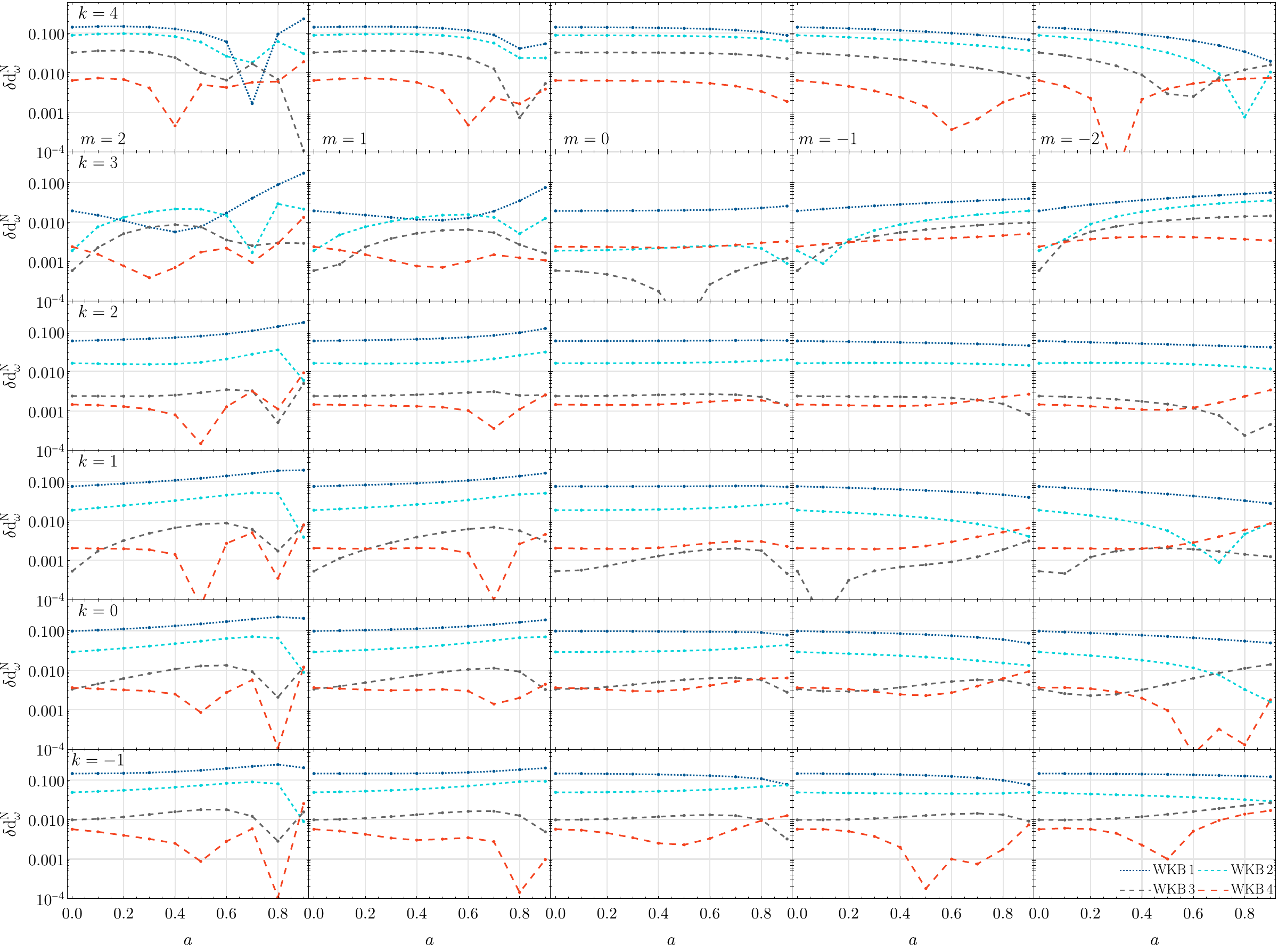}
    \caption{Comparison of the linear coefficients $d^N_\om$ for $(2,m,0)$ modes computed with $N$-th order WKB and Leaver's method at different spins, $a$. 
    We plot the relative errors $\delta d^N_\omega$ defined in Eq.~\eqref{rel_err_d}. Different columns correspond to different values of $m$, and different rows correspond to different values of $k$. Colors indicates WKB orders, with the same conventions used in Fig.~\ref{fig:kerrgrwkb}. \label{fig:beyondTeukolskyrj}}   
\end{figure*}

The WKB method has been applied in the past to gravitational QNMs of Kerr BHs~\cite{Seidel:1989bp,Kokkotas:1991vz} (although the results are limited to just a few modes and low WKB orders) and to scalar perturbations of rotating regular spacetimes that can mimic Kerr BHs~\cite{Franzin:2022iai}.  

In Ref.~\cite{Seidel:1989bp}, Seidel and Iyer computed gravitational QNMs starting from the purely real Chandrasekhar-Detweiler potential,  that can be found by a suitable transformation of the Teukolsky equation~\cite{Chandrasekhar:1976zz}.
They expanded the separation constant at sixth order in the spin and evaluated QNMs with third-order WKB for different values of $\ell,m,n$, using a Pad\'e resummation for the spin expansion. Their tabulated results are available for $\ell=2$, $m=0,1$.
Kokkotas~\cite{Kokkotas:1991vz} applied the WKB expansion directly to the complex-valued Teukolsky potential, finding good agreement with the results available at the time. That analysis was also made at third order in the WKB approximation, and by expanding the separation constant up to sixth order in the spin. Only the $\ell=2$, $m=0$ results were tabulated.
The accuracy of these calculations is limited by the spin expansion of the separation constant $B_{\ell m}$, which leads to relatively large errors at high spins.
In this work, we use the separation constants provided by the Black Hole Perturbation Toolkit~\cite{BHPToolkit}, which can be computed to arbitrary precision at every value of the spin.

It is convenient to define the relative difference between QNMs computed with the WKB expansion (from first to fourth order) and Leaver's method,
\begin{equation}\label{rel_err_kerr}
    \de \om_{\ell m n} = \left| \frac{\om_\note{WKB}}{\om_\note{Leaver}} - 1 \right|\,.
\end{equation}

In Fig.~\ref{fig:kerrgrwkb}, we plot $\de \om_{\ell m n}$ for a wide range of modes: $\ell =2$, $m=[-\ell,\,\ell]$ and $n=[0,\,2]$. The QNMs are computed at values of the spin $a\in [0,\,\dots,\,0.9]$ in steps of $\de a=0.1$. Each step is marked with a dot along the curves. For completeness, in Appendix~\ref{app:HM_WKB} we present similar plots for $\ell =3,\,4$. The QNM calculations using the WKB method have been performed with \texttt{Mathematica}'s built-in function \texttt{FindRoot} by imposing machine precision accuracy. These settings were used in all the calculations reported in the paper.

In general, the WKB approximation works best for large $\ell$ and low $n$ ($n \lesssim \ell$). The lowest relative errors are for the $(\ell,m,n) = (4,\,m,\,0)$ modes, followed by the $(3,\,m,\,0)$ modes, where higher WKB orders lead to a substantial improvement in the accuracy of the QNM frequencies. For the $(2,m,0)$ modes, the accuracy increases monotonically with the WKB order only when $m\leq0$. For the $(2,\,2,\,0)$ and $(2,\,1,\,0)$ modes the improvement typically saturates at third order, and a fourth-order expansion does not bring a substantial increase in accuracy.
As expected, the relative errors $\de \om_{\ell m n}$ worsen for all overtones ($n\geq1$). 
Increasing the WKB order helps at moderate spins, but we find the quite general trend that increasing the value of $a$ induces a saturation of the results at WKB order $\geq2$.

In Fig.~\ref{fig:kerrgrwkb}, we also compare our results with the third-order WKB results by Kokkotas for the $(2,0,0)$ mode~\cite{Kokkotas:1991vz}, and with the third-WKB order results of Seidel-Iyer~\cite{Seidel:1989bp}, available for $\ell=2$, $m=2,1$ and $n=0$. 
We find near-perfect agreement with the results found by Kokkotas. The observed differences with respect to the Seidel-Iyer calculation are understandable because their approach was totally different, as we discussed above.

\section{Results for beyond-Kerr}\label{sec:res_bGR}

In this section, we present WKB calculations of beyond-GR QNM frequency corrections, first in the ``theory agnostic'' modified Teukolsky framework, and then in the specific case of HDG theories.

\subsection{Modified Teukolsky Equation}

The computation of Kerr QNMs with the WKB method allows us to estimate the linear coefficients $d_\om^N$ through Eq.~\eqref{eq:linear_coeff_WKB}. 
In Fig.~\ref{fig:beyondTeukolskyrj} we plot the relative difference
\begin{equation}\label{rel_err_d}
    \de d_\om^N = \left| \frac{d_\om^N}{d_\om} - 1 \right| \,,
\end{equation}
where we recall that $N$ stands for the WKB order and $d_\om$ are the same coefficients computed with the Leaver method~\cite{Cano:2024jkd}. 
We show results for the $(2,m,0)$ modes with $m=[-2,\dots,2]$ and for radial power modification spanning the values $k=[-1,\dots,4]$. For all values of $k$, the accuracy increases with the WKB order, as expected. 
We have checked other combinations of $\ell$ and $n$ and the results look qualitatively similar.

\subsection{Higher-Derivative Gravity}

\begin{figure}[ht!]
    \centering
     \includegraphics[width=\columnwidth]{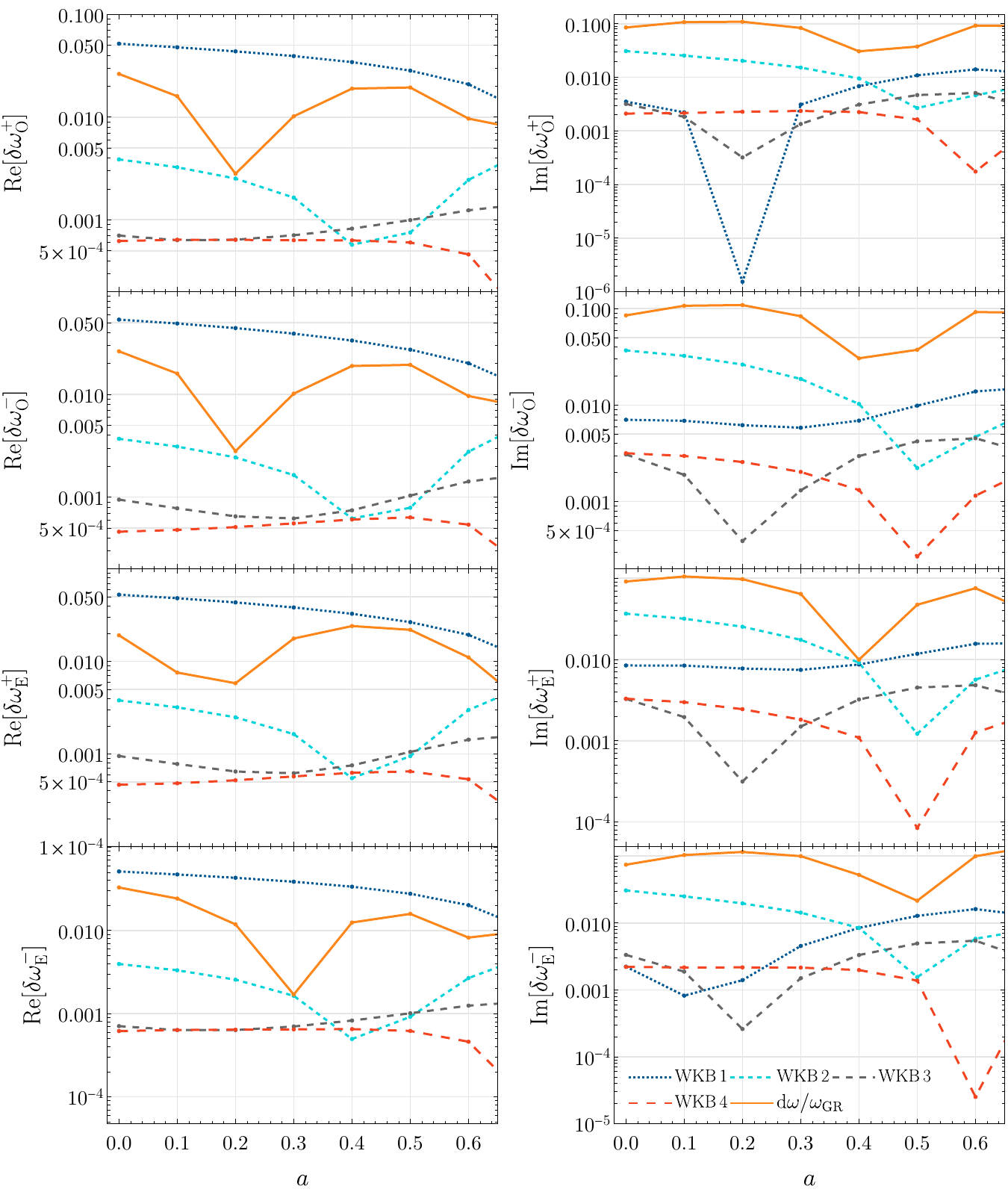}
    \caption{Comparison of QNM relative differences $\de \om^\pm_\note{e/o}$ in a HDG theory with coupling $\la_\note{eff}=0.1$, as defined in Eq.~\eqref{rel_err_omega_hdg}. The left (right) panels refer to the real (imaginary) part of the $(2,2,0)$ mode. Each row selects a combination of polarization ($\pm$) and parity of the theory (even/odd). Orange solid lines show the relative difference between modes in GR and HDG.} 
  \label{fig:linear_WKB_HDG_la=0.1}
\end{figure}

Finally, we consider HDG theories as an example of beyond-GR theories. These theories of gravity include higher-order curvature terms in the gravitational action~\cite{Endlich:2017tqa,Cano:2019ore,Ruhdorfer:2019qmk}. In Ref.~\cite{Cano:2024ezp} it was shown that the perturbation equations of rotating BHs in HDG can be recast in the form of Eq.~\eqref{eq:mod_teukolsky}, where the only non-vanishing values of $\al^{(k)}$ in the modified potential are $k = k^\note{HDG}=[-2,0,1,2]$. As discussed in~\cite{Cano:2023tmv}, the effect of these modifications is to break isospectrality between even $(+)$ and odd $(-)$ perturbations, leading to a modified potential of the form
\begin{equation}
    \de V^\pm = \la_\note{eff} \sum_{k\in k^\note{HDG}} \al_\pm^{(k)} \left( \frac{r}{r_+} \right)^k \,,
\end{equation}
where we have factored out the normalized coupling constant $\la_\note{eff}$ for the selected HDG theory.

First, we compare the $(2,2,0)$ and the $(2,1,0)$, $(\pm)$ polarizations of the even and odd $(\note{E/O})$ HDG modifications with $\la_\note{eff} = 0.1$. The modes are computed by using either the WKB linearized parametrization through the coefficients $d_\om^N$, labeled as $\om^{\pm,\note{WKB}}_\note{e/o}$, or the continued fraction (Leaver) linearized parametrization through the coefficients $d_\om$, labeled as $\om^{\pm,\note{Leaver}}_\note{e/o}$.
Again, we present relative differences between QNMs as
\begin{equation}\label{rel_err_omega_hdg}
    \de \om^\pm_\note{e/o} = \left| \frac{\om^{\pm,\note{WKB}}_\note{e/o}}{\om^{\pm,\note{Leaver}}_\note{e/o}} - 1 \right| \,.
\end{equation}
We carry out the comparison only up to $a=0.65$, the (conservative) maximum value of the spin such that the modifications in the potential $\de V^\pm$ are reliable. The results are displayed in Fig.~\ref{fig:linear_WKB_HDG_la=0.1}. It is worth noting that going from the third to the fourth order, the gain in the relative difference is mild or absent. This is in contrast to the behavior of the single coefficients shown in Fig.~\ref{fig:beyondTeukolskyrj}.

In Fig.~\ref{fig:linear_WKB_HDG_la=0.1} we also plot the relative difference between HDG and GR modes, which by definition is given by $\la_\note{eff} d_\om / \om^\note{GR}$. It is worth noting that for the chosen coupling, the deviation from GR is always much larger than the error introduced by the WKB approximation (with $N\geq 2$) with respect to the ``true'' value obtained with continued fractions. In Fig.~\ref{fig:linear_WKB_HDG} of Appendix~\ref{app:WKBvsLeaver_for_HDG} we reach similar conclusions for larger coupling ($\la_\note{eff} = 1$).

Note that a $1\%$ relative error in the real part and a $10\%$ error in the imaginary part are comparable to the accuracy of the ringdown tests performed with GW250114~\cite{LIGOScientific:2025rid,LIGOScientific:2025obp}. This event yields agnostic constraints on the fundamental frequency of $\de \mathrm{Re}(\om) / \mathrm{Re}(\om)  = 0.029$ and $\de \mathrm{Im}(\om) / \mathrm{Im}(\om) = 0.17$ if we consider the public results of the Bayesian analysis performed with a sum of damped sinusoids starting $10M_f$ after the merger, where $M_f=68.1 M_\odot$ is the maximum likelihood final mass from the full inspiral-merger-ringdown analysis.

\begin{figure}
    \centering
    \includegraphics[width=\linewidth]
    {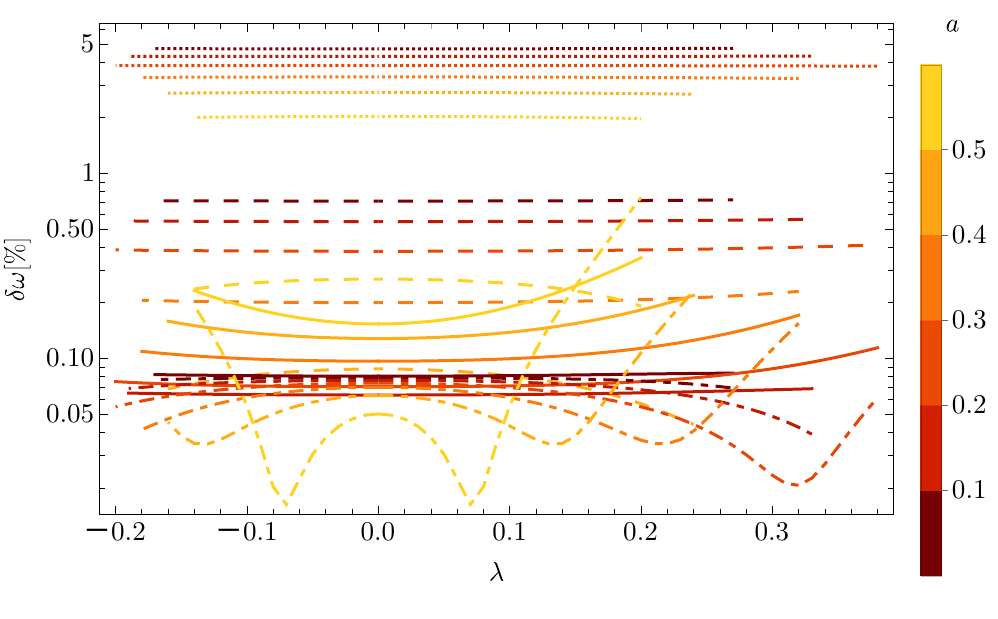}
    \caption{Percentage relative error on the $\ell=m=2$ fundamental QNM frequencies computed with continued fractions and with the $N$-th order WKB approximation, for different values of $\la$ and $a$. Dotted lines correspond to $N=1$, dashed lines to $N=2$, solid lines to $N=3$, and dot-dashed lines to $N=4$. Each curve spans values of $\la$ for which the relative difference between the linearized QNM calculation and the continued fraction calculation is less than $5\%$.}
    \label{fig:WKB_vs_CF_comparison}
\end{figure}

As we discussed in Sec.~\ref{sec:WKB_theory}, the WKB method does not require linearization around GR, as long as the potential has a single peak and two turning points with suitable boundary conditions. Hence, we can ask how accurate the standard higher-order WKB approximation is with respect to continued fraction calculations. We proceed as follows. We select a target value for $\la_\note{eff}$ and $a$. We first compute the QNMs for $a=0$ and $\la_\note{eff}=0$, using the analytic prediction for $\overline{r}_+$ given in Eq.~\eqref{eq:max_r} and the numerical GR prediction for $\om_N$. Then, we start increasing $\la_\note{eff}$ by little steps, at each step solving numerically for $\overline{r}$ and $\om$, using the results of the previous step as guesses, until the targeted value of $\la_\note{eff}$ is obtained. Then, by fixing $\la_\note{eff}$, we start exploring different values of the spin.

A summary of the results is reported in Fig.~\ref{fig:WKB_vs_CF_comparison}. 
Here, we present the relative errors $\de\om$ between the exact $\ell=m=2$ fundamental mode frequencies obtained by Leaver's method and those obtained at different WKB orders for various values of $\lambda$ and $a$.
We truncate the lines at values of $\lambda$ for which the relative errors between the linear shift and the Leaver prediction correspond to 5\,\%. 
Increasing the WKB order clearly reduces the relative errors for the first three WKB orders. 
In contrast, from the fourth WKB order onwards we observe oscillatory patterns, and the approximation can perform either better or worse than the previous order depending on the value of $\lambda$. 
Overall, the WKB results typically improve with increasing spin $a$, but the range of validity of the linear shift (as shown by the truncation of the lines in the plot) also gets smaller.

\section{Conclusions}\label{sec:conc}

The accurate computation of QNMs for rotating BHs beyond GR is an important goal of BH spectroscopy. 
While the higher-order WKB method is a widely used and popular tool for computing QNMs for nonrotating BHs~\cite{Schutz:1985km,Iyer:1986np,Iyer:1986nq,Kokkotas:1988fm,Konoplya:2003ii,Matyjasek:2017psv}, its application to rotating BHs has received relatively little attention~\cite{Seidel:1989bp,Kokkotas:1991vz}. 
This is due in part to the scarcity of separable master equations for perturbations of rotating BHs in theories beyond GR. 
Furthermore, the technical complexity of applying the higher-order WKB method to such cases is increased by the need for careful treatment of complex-valued potentials and nontrivial separation constants. 
In this work, we have applied the higher-order WKB approximation to several nontrivial cases. 

In Sec.~\ref{sec:res_GR}, we have revisited early results for Kerr BH in GR by systematically comparing the WKB-approximated QNMs and their ``exact'' values obtained from a Leaver continued fraction code. 
We find an improvement in convergence against the results presented in Ref.~\cite{Seidel:1989bp}, especially at large spins, mainly because those authors performed a sixth-order Taylor expansion in the spin to obtain the separation constant. Our results also agree with those reported in~\cite{Kokkotas:1991vz}. Furthermore, we studied all the modes with $\ell = [2,4]$, $m = [-\ell , \ell]$, and $n=[0,2]$ using WKB expansions ranging from first to fourth order. 
We have found that QNMs with low overtone numbers $n$ and large harmonic indices display the best agreement with continued fraction results. 
As expected, the accuracy of the calculation gets worse as $n$ grows and as $\ell$ decreases. We also noticed that the WKB expansion seems to have better convergence properties for negative values of $m$. 
For the fundamental mode, the accuracy improves as the WKB order increases, whereas for overtones this trend is not as clear, and relative errors tend to saturate at second order.  

In Sec.~\ref{sec:res_bGR}, we have extended this comparison to the recently introduced theory-agnostic parametrization of the modified Teukolsky equation~\cite{Cano:2024jkd} and to HDG theories~\cite{Cano:2024ezp}. 
We have found that the relative difference of the linearized coefficients computed with the WKB method against continued fractions results follows trends similar to the Kerr QNMs, with an overall uniform behavior across spin values and improved accuracy at higher WKB orders. 

A similar trend was found for QNMs in HDG theories. Even for relatively small values of the coupling, the relative error of this calculation is smaller than the relative difference between GR and HDG, provided that we push the WKB expansion at least to second order. 
Moreover, these relative differences are lower than the measurement errors on the fundamental mode for GW250114, the event with the highest ringdown signal-to-noise ratio  detected so far~\cite{LIGOScientific:2025rid,LIGOScientific:2025obp}. 

This implies that an agnostic parametrization using WKB predictions might be accurate enough to perform tests of GR with current observations, and to leverage the free ``null test'' parameters included in these tests.
However, it is not clear whether the complex-valued nature of the effective potential allows for a simple interpretation as in the nonrotating case, where QNMs are associated to the local properties (height and curvature) of the potential close to its maximum~\cite{Volkel:2022khh,Thomopoulos:2025nuf}. 
The identification of the light ring as the approximate location of the effective potential is, in general, also different. 
In the higher-order WKB method, the extremum of the effective potential is determined by the complex-valued roots of $\text{d}Q(r, \omega)/\text{d}r_*$, which also make the height and the second-derivative complex-valued, thus doubling the free parameters compared to the nonrotating case. 
An intriguing direction for future work is to explore whether it is possible to find a real-valued, isospectral effective potential whose local properties are determined from the QNMs. A possible approach would be to transform the modified Teukolsky equation in a way analogous to the Chandrasekhar-Detweiler transformation~\cite{Chandrasekhar:1976zz}. Another possibility would be to extend the strategy of Refs.~\cite{Albuquerque:2024xol,Albuquerque:2024cwl}, where a frequency-dependent potential is mapped to a real-valued barrier for the calculation of transmission coefficients and bound states using WKB theory.

It is also important to quantify the signal-to-noise ratio at which the higher-order WKB approximation may introduce biases in constraints on beyond-GR theories, due to small but finite QNM inaccuracies. 
This is relevant because next-generation detectors will deliver very small statistical errors on QNMs, potentially below the systematic errors associated with the WKB method. 
A first step in this direction would be to use linear-signal analysis to estimate parameter biases (e.g., in mass and spin), as in Refs.~\cite{Capuano:2025kkl,Volkel:2025jdx}.

\acknowledgments

R.T.~and N.F.~would like to thank the Johns Hopkins University for hospitality during the early stages of this work.
N.F.~acknowledges funding from the FCT grant agreement 2023.06263.CEECIND/CP2830/CT0004 and support to the Center for Astrophysics and Gravitation (CENTRA/IST/ULisboa) through FCT grant No.~UID/PRR/00099/2025 and grant No.~UID/00099/2025.
S.\,H.\,V. acknowledges funding from the Deutsche Forschungsgemeinschaft (DFG): Project No. 386119226.
E.\,B.~is supported by NSF Grants No.~AST-2307146, No.~PHY-2513337, No.~PHY-090003, and No.~PHY-20043, by NASA Grant No.~21-ATP21-0010, by John Templeton Foundation Grant No.~62840, by the Simons Foundation [MPS-SIP-00001698, E.B.], by the Simons Foundation International [SFI-MPS-BH-00012593-02], and by Italian Ministry of Foreign Affairs and International Cooperation Grant No.~PGR01167.
This work was carried out at the Advanced Research Computing at Hopkins (ARCH) core facility (\url{https://www.arch.jhu.edu/}), which is supported by the NSF Grant No.~OAC-1920103.
\appendix

\section{Higher modes QNM comparison}\label{app:HM_WKB}

\begin{figure*}
    \centering
    \includegraphics[width=0.9\linewidth]{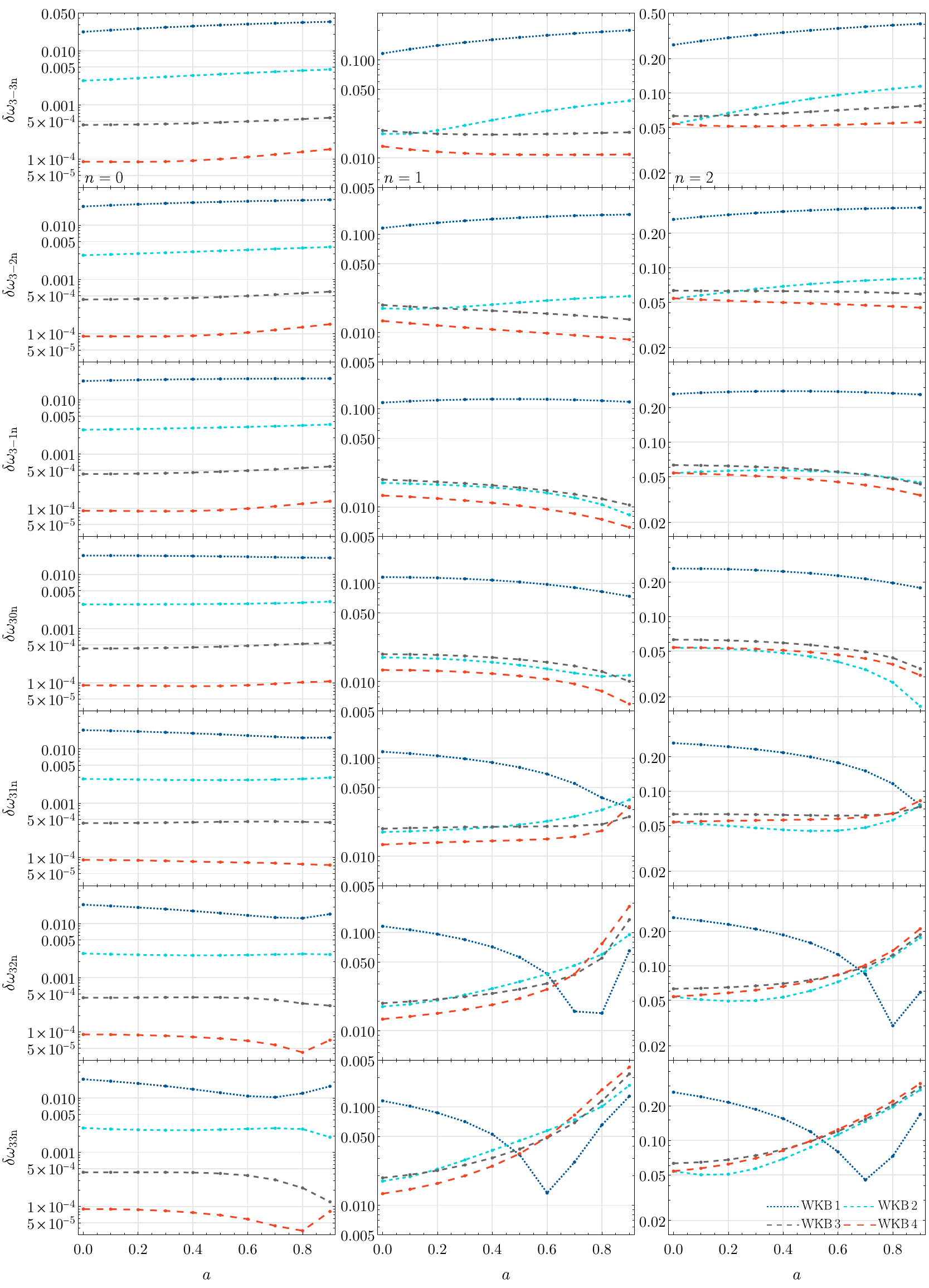}
    \caption{Same as Fig.~\ref{fig:kerrgrwkb}, but for $\ell = 3$. \label{fig:Kerr_l3}}
\end{figure*}

\begin{figure*}
    \centering
    \includegraphics[width=0.7\linewidth]{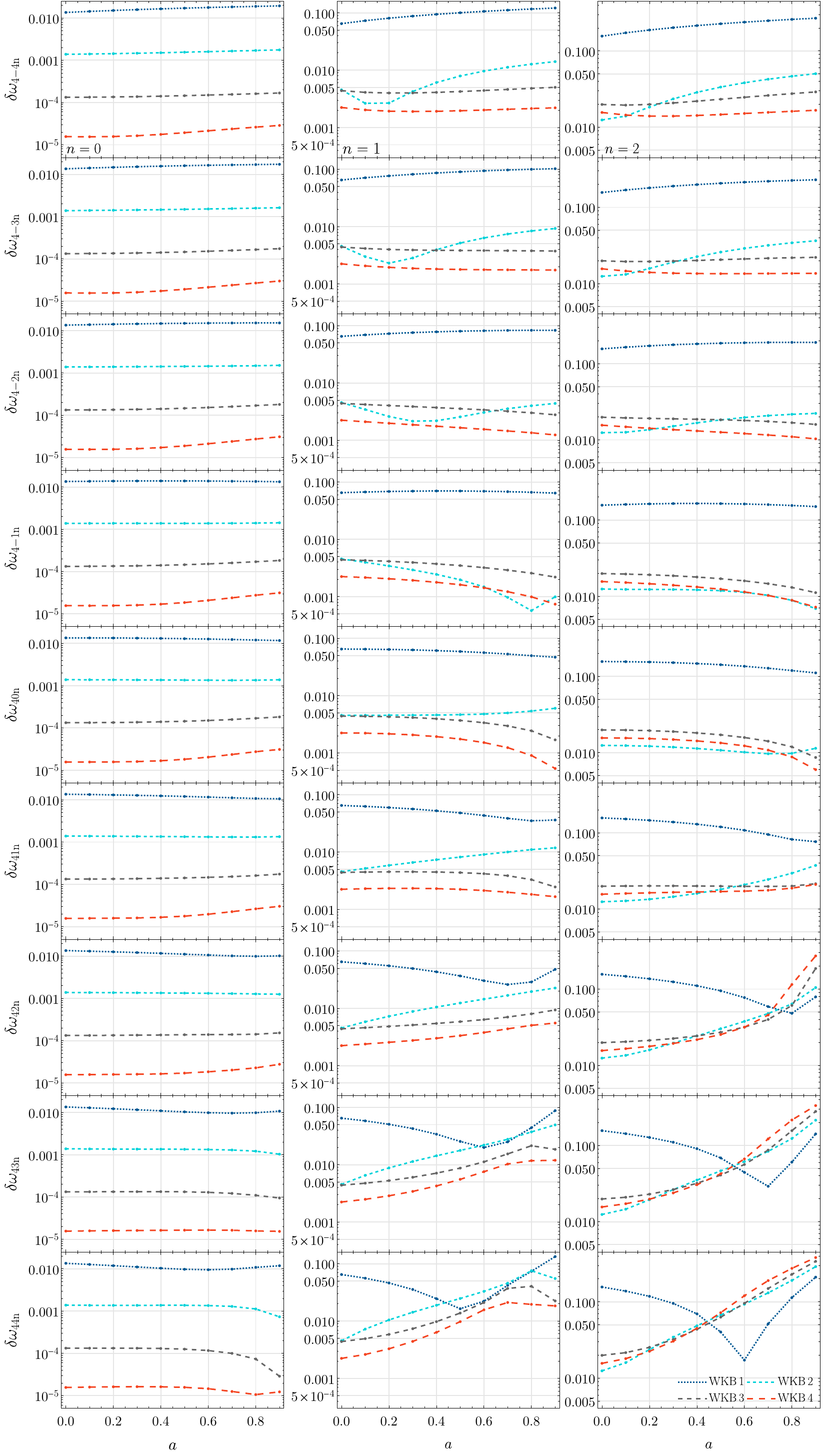}
    \caption{Same as Fig.~\ref{fig:kerrgrwkb}, but for $\ell = 4$. \label{fig:Kerr_l4}}    
\end{figure*}

In this Appendix, we report the relative difference between the WKB approximation and Leaver's method for Kerr QNMs with $\ell=3$ (Fig.~\ref{fig:Kerr_l3}) and $\ell = 4$ (Fig.~\ref{fig:Kerr_l4}).

\section{Comparison of WKB results for higher-derivative gravity}\label{app:WKBvsLeaver_for_HDG}

\begin{figure}
    \centering
     \includegraphics[width=\hsize]{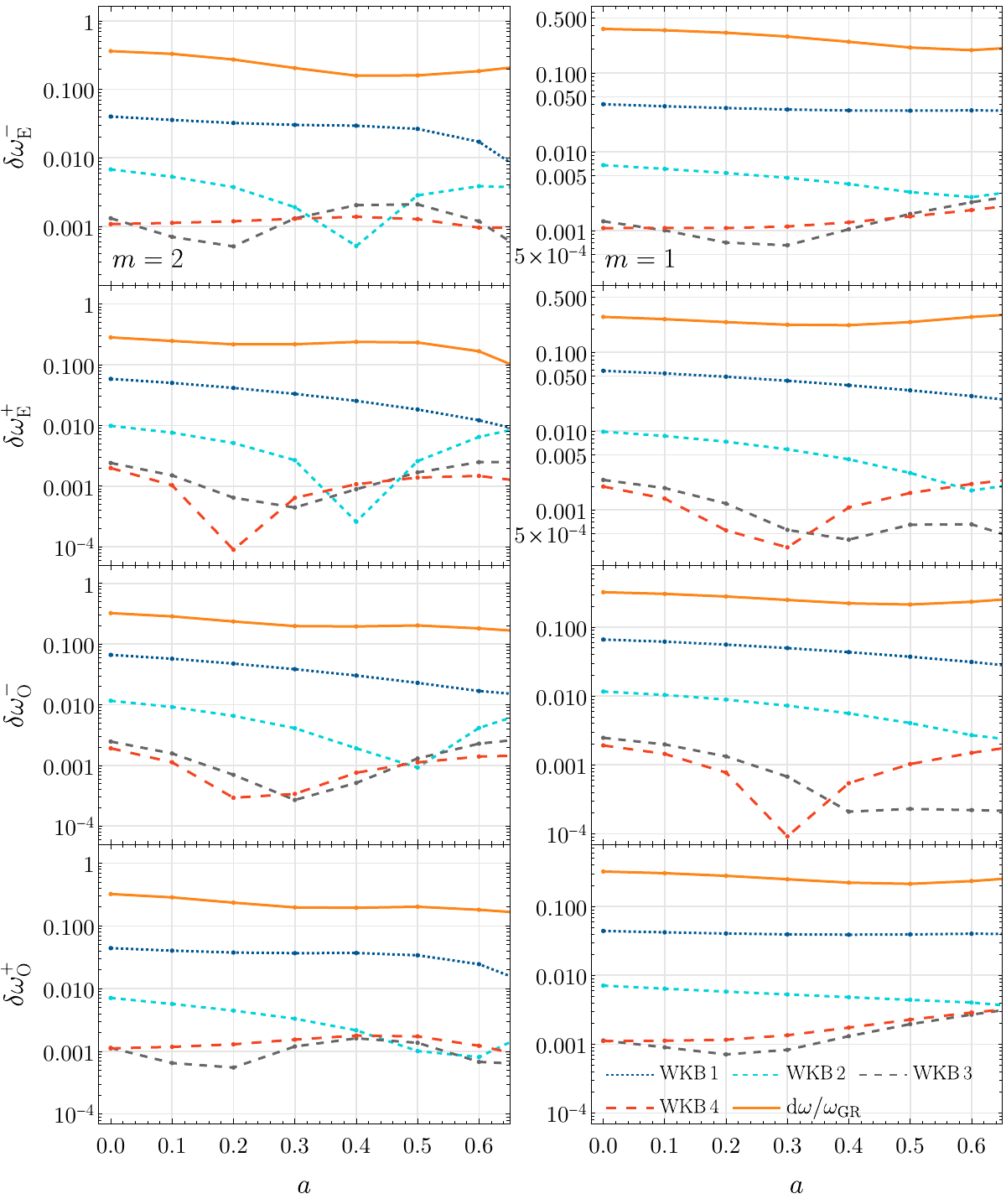}
    \caption{Same as Fig.~\ref{fig:linear_WKB_HDG_la=0.1}, but for $\la_\note{eff} = 1$. \label{fig:linear_WKB_HDG}}
\end{figure}

\begin{figure}
    \centering
     \includegraphics[width=\hsize]{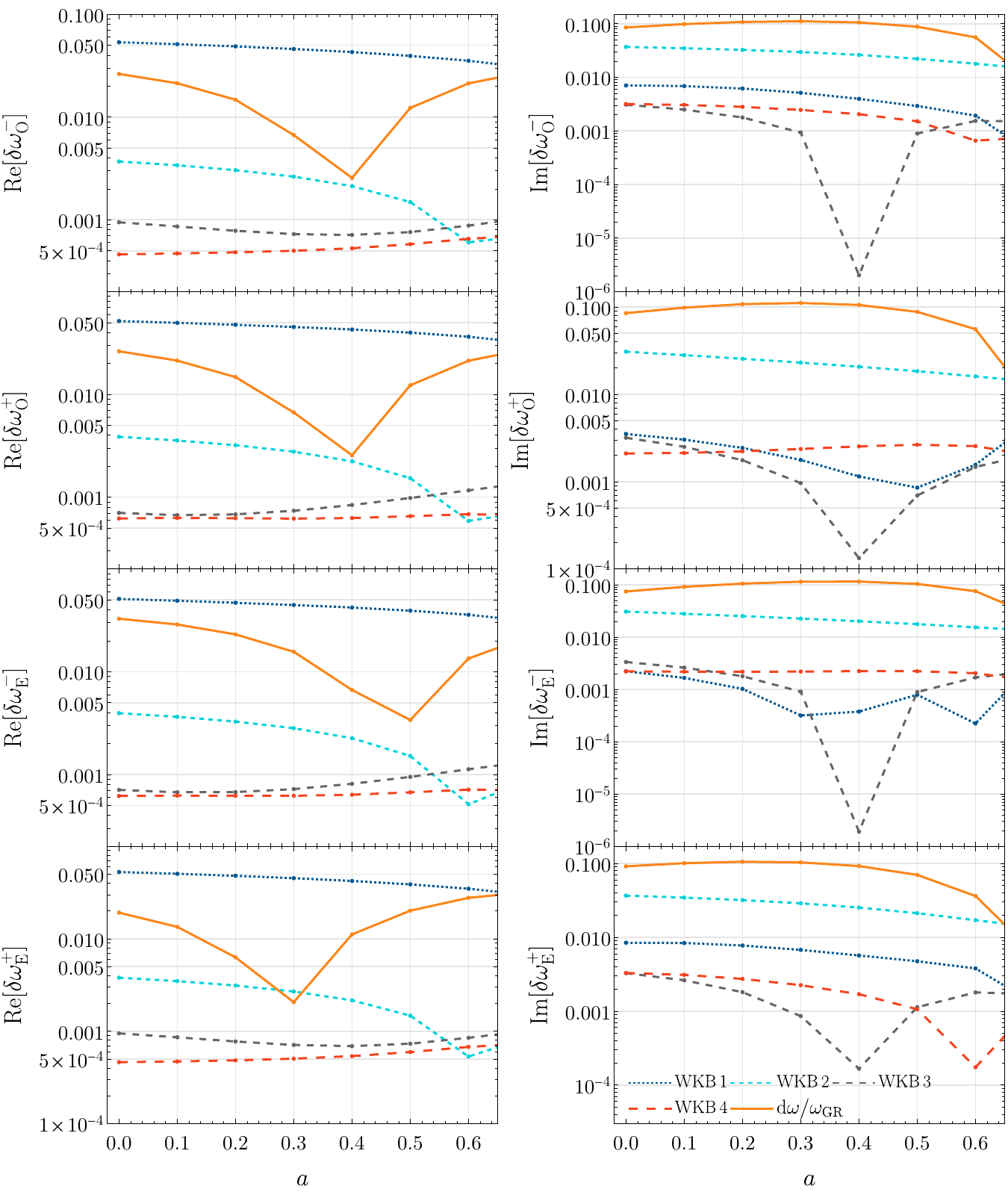}
    \caption{Same as Fig.~\ref{fig:linear_WKB_HDG_la=0.1}, but for $(\ell,\,m,\,n)=(2,\,1,\,0)$. 
    \label{fig:linear_WKB_HDGla=0.1-210}}
\end{figure}

In this Appendix, we provide further comparisons between the WKB approximation and continued fraction calculations of QNMs in HDG. 
In Fig.~\ref{fig:linear_WKB_HDG} we report the same comparison as in Fig.~\ref{fig:linear_WKB_HDG_la=0.1}, but with a different choice of coupling (here, $\la_\note{eff} = 1$). For this value of the coupling, the relative difference between GR and HDG is much larger than any systematics introduced by the WKB approximation. Finally, in Fig.~\ref{fig:linear_WKB_HDGla=0.1-210} we consider the same coupling as in the main text ($\la_\note{eff} = 0.1$), but we focus on the $(2,1,0)$ QNM.

\bibliography{literature}

\end{document}